\documentclass{aa}  

\newcommand{\todo}[1]{}
\renewcommand{\todo}[1]{{\color{red} TODO: {#1}}}

\usepackage{lipsum,graphicx,multicol}
\usepackage{CJKutf8}
\usepackage{multirow}
\usepackage[switch]{lineno}
\usepackage{tabularx}
\usepackage{graphicx}
\usepackage{gensymb}
\usepackage{tabularray}
\usepackage{tablefootnote}
\usepackage{subcaption}
\usepackage{txfonts}
\usepackage{natbib}
\usepackage[colorlinks, citecolor=blue]{hyperref}
\bibpunct{(}{)}{;}{a}{}{,} 
\begin{document} 

   \title{The evolution of extragalactic peaked-spectrum sources down to 54 megahertz} 

   \subtitle{}

   \author{Sai Zhai (\begin{CJK*}{UTF8}{gbsn}翟赛\end{CJK*})\inst{1}
            \and
            Anniek J. Gloudemans \inst{2}
            \and
          G\"ulay G\"urkan \inst{3,4}
          \and
          Femke J. Ballieux \inst{1}
          \and
            Martin J. Hardcastle \inst{3}
            \and
            \newline
            Francesco De Gasperin \inst{5}
          \and
          Huub  J.A. Röttgering\inst{1}
          }
   \institute{Leiden Observatory, Leiden University, P.O. Box 9513, 2300 RA Leiden,
The Netherlands\\
              \email{szhai@strw.leidenuniv.nl}
        \and 
        NSF NOIRLab, Gemini Observatory, 670 N A'ohoku Place Hilo, HI 96720, USA
         \and
        Centre for Astrophysics Research, University of Hertfordshire, College Lane, Hatfield AL10 9AB, UK
        \and
        CSIRO Space and Astronomy, ATNF, PO Box 1130, Bentley, WA 6102, Australia
        \and
        INAF - Istituto di Radioastronomia, via P. Gobetti 101, 40129, Bologna, Italy
         \\
            }

   \date{\textit{Received 2024 November 13; accepted xx}}

  \abstract
  {Peaked-spectrum (PS) sources, known for their distinct peaked radio spectra, represent a type of radio-loud active galactic nuclei (AGN). Among these, megahertz-peaked spectrum (MPS) sources, which exhibit a spectral peak at a frequency of a hundred megahertz, have emerged as a potential tool for identifying high-redshift candidates. However, the potential evolutionary link between the fraction of these sources and redshift remains unclear and requires further investigation. The recent, high sensitivity Low Frequency Array (LOFAR) surveys enable statistical studies of these objects to ultra-low frequencies (< 150 MHz). In this study, we first use the multiradio data to investigate the evolution of spectral index with redshift for 1,187 quasars from the SDSS 16th quasar catalog. For each quasar, we analyze available data from the LOFAR Low Band Antenna (LBA) at 54 MHz, High Band Antenna (HBA) at 144 MHz, and the Very Large Array (VLA) the Faint Images of the Radio Sky at Twenty cm (FIRST) at 1.4 GHz. We measure the spectral index ($\alpha^{144}_{54}$ and $\alpha^{1400}_{144}$) and find no significant change in their median values with the redshift. Extended sources have steeper spectral indices than compact sources, which is consistent with previous findings. Based on the spectral indices information, we identify MPS sources using these criteria: $\rm \alpha^{144}_{54} >= 0.1$ and $\rm \alpha^{1400}_{144} < 0$, and analyze their properties. We find that the fraction of MPS sources is constant with the redshift ($0.1-4.8$), bolometric luminosity ($\rm 10^{44}-10^{48} erg/s$), and supermassive black hole mass ($\rm 10^{7}-10^{10.5} M_{\odot}$), which suggests that MPS sources have relatively stable physical conditions or formation mechanisms across various evolutionary stages and environments.}


   \maketitle
%

\section{Introduction}

    Quasars are among the most luminous and energetic objects in the universe (e.g., \citealt{1963Natur.197.1040S,1999PASP..111..661S}). These active galactic nuclei (AGN) are powered by supermassive black holes (SMBHs) at the center of galaxies through the accretion of large amounts of material (e.g., \citealt{1984ARA&A..22..471R,1995ARA&A..33..581K}). Their extreme luminosities allow us to detect them at great distances, thereby enabling the study of the high redshift universe, which offers insights into various subsets of AGN, including the peaked-spectrum (PS) sources (e.g., \citealt{2006ARA&A..44..415F,10.1093/mnras/stv681,2020ApJ...897L..14Y,2021ApJ...907L...1W}).

    PS sources are characterized by their peaked radio spectra. Based on their peak frequency ($\nu_{o}$), they are divided into high-frequency peaked (HFP) sources that peak above 5 GHz, gigahertz-peaked spectrum (GPS) sources that peak at 1-5 GHz, compact steep spectrum (CSS) sources that peak below 500 MHz, and megahertz-peaked spectrum (MPS) sources that peak below 1 GHz in the observed frame \citep{1990A&A...231..333F, 1991ApJ...380...66O,2000A&A...363..887D,2016MNRAS.459.2455C,2017ApJ...836..174C, 2024A&A...689A.264B}. These sources exhibit small linear sizes: HFP and GPS sources are smaller than 1 kpc, while CSS sources range from 1-20 kpc \citep{1998PASP..110..493O}. Previous work also found larger sources tend to have lower
    peak frequency \citep{1997AJ....113..148O,2000MNRAS.319..445S}. 


Despite extensive research, the reason behind the turnover in their radio spectra remains unclear. Two models—the "youth" and "frustration" models—have been proposed to explain these turnovers. The "youth" model suggests that these sources follow an evolutionary sequence, with HFP sources gradually maturing into GPS sources, then CSS sources, and ultimately evolving into FR I and FR II radio galaxies \citep{2002NewAR..46..307M, 2010MNRAS.408.2261K}. In contrast, the "frustration" model posits that these sources are not young but their small size is attributed to confinement by the surrounding dense, hot nuclear plasma, which slows the expansion of the sources \citep{1984AJ.....89....5V}. It is also possible that both mechanisms contribute to the turnover of PS sources \citep{2012ApJ...760...77A,2017ApJ...836..174C}. It is also possible that both Synchrotron Self-Absorption (SSA, suggested by the "youth" model) and Free-free absorption (FFA, suggested by the "frustration" model) contribute to these spectral turnovers \citep{1997ApJ...485..112B,2014ApJ...780..178M,2019A&A...628A..56K}. Typically, the "youth" model predicts a spectral index near 2.5, while the "frustration" model suggests a spectral index closer to 2.1, although external ionized gas could produce even steeper values \citep{1997ApJ...485..112B, 2015ApJ...809..168C, 2019A&A...628A..56K}. Thus, measuring the spectral index below the peak frequency is crucial for distinguishing between these two mechanisms.
    
    Understanding these spectral properties has broader implications, especially in the search for high redshift AGN. In recent years, MPS sources have been used to select high redshift AGN \citep{2004NewAR..48.1157F,10.1093/mnras/stv681}. Earlier methods focused on using ultra-steep spectrum (USS) sources, with a spectral index steeper than $-1.3$, to locate high redshift sources. This approach was based on the observation that high redshift radio galaxies (HzRGs) tend to have steeper radio spectra compared to their local counterparts \citep{2000A&AS..143..303D, 2008A&ARv..15...67M}. This "$\alpha-z$" relationship, which links spectral steepness to redshift, led researchers to explore MPS sources as a new method for identifying high redshift AGN \citep{2009AJ....137.4846O,2012MNRAS.420.2644K}. This approach is based on the idea that observed MPS sources represent a combination of high redshift GPS and CSS sources, where the peak frequency is shifted due to cosmological evolution \citep{2017ApJ...836..174C}. From this perspective, \cite{2004NewAR..48.1157F} proposed that compact MPS sources (on the order of tens of milliarcseconds) could be an effective tool for finding high redshift AGN. This method was successfully validated by \cite{10.1093/mnras/stv681}, who found 33 MPS sources at high redshift (approximately $z \sim 2$).

    The advent of extremely low-frequency observations with the Low Frequency Array (LOFAR; \citealt{2013A&A...556A...2V}) 
    makes it possible to find more MPS sources and study the physical mechanisms behind PS sources \citep{2022A&A...668A.186S,2024A&A...689A.264B}. The LOFAR Low Band Antenna Sky Survey (LoLSS) data release 1 (DR1) offers high sensitivity (1-2 mJy/beam) and high resolution (15$''$) observations, covering 650 $\rm deg^{2}$ at 42-66 MHz \citep{2023A&A...673A.165D}. Additionally, the LOFAR Two-metre Sky Survey (LoTSS) DR2 provides even higher sensitivity (<100 $\mu$Jy/beam) and resolution (6$''$), covering 5635 $\rm deg^{2}$ at 120-168 MHz \citep{2022A&A...659A...1S}. Complementing LOFAR data with other large sky radio surveys enables us to expand the study of peaked spectrum sources to frequencies below 100 MHz.

    
    In this work, we use multiple radio surveys to study the $"\alpha-z"$ relationship of quasars at lower frequencies. Among them, we selected MPS sources to investigate their occurrence from local to high redshift ($z$=0.1-4.8) and study their evolutionary trends. In Section 2, we discuss the selection process for the quasar samples and MPS sources. In Section 3, we present the spectral index evolution of quasars and the fractions of MPS sources as a function of redshift, bolometric luminosity, and SMBH mass. In Section 4, we focus on interpreting these results, specifically examining how these dependencies reflect the physical processes influencing quasar evolution and MPS source distributions. The summary is presented in Section 5. We use the following cosmological parameters: $H_0 = 67.8 \text{ km s}^{-1} \text{ Mpc}^{-1}$, $\Omega_m = 0.308$, and $\Omega_\Lambda = 0.692$ \citep{2016A&A...594A..13P}. The spectral index used in this work is defined as $S_{\nu} \propto \nu^{\alpha}$, where $S_{\nu}$ is the flux density, $\nu$ is the frequency and $\alpha$ is the spectral index.


\section{Data}

\begin{table} 
\centering
\caption{Sample selection criteria}
\begin{tblr}{
  width = \linewidth,
  colspec = {Q[138]Q[500]Q[280]},
  cells = {c},
  cell{2}{1} = {r=4}{},
  vline{2-3} = {1-13}{},
  vline{3} = {3-5}{},
  hline{1,13} = {-}{0.16em},
  hline{2} = {-}{0.16em},
  hline{6,7,8,11,12} = {-}{0.16em},
  hline{9-10} = {-}{},
  hline{3-5} = {2-3}{}
}\label{table1}
\textbf{\textbf{Selection step~}} & \textbf{\textbf{Selection criterion}}                                 & \textbf{\textbf{Number of sources}}     \\
0                                 & The LoTSS DR2 catalog with optical identifications                                       & 4,116,934 radio sources                 \\
                                  & The SDSS 16th quasar catalog                                                      & 750,414 quasars                         \\
                                  & The~LoLSS~DR1 catalog~                                                & 42,463 radio sources                    \\
                                  & The FIRST catalog                                                     & 946,432 sources                         \\
1                                 & LoTSS quasar samples (Find the SDSS counterparts of LoTSS samples within 5$''$)  & 64,464 quasars \\
2 & LoTSS quasar samples with both LoLSS (within 5$''$) and FIRST (2$''$) detections &
1,187 quasars                            \\
2.1                              & Separate the extended and~compact sources                             & 232 extended sources, 955 compact sources                    \\
2.2                               & Separate the radio and~optical quasar samples                         & 1037 optical quasars,~150 radio quasars \\
2.3   &$\alpha_{54}^{144}  \geq 0.1 $ and $\alpha_{144}^{1400}<0$  & 61 MPS sources \\
3 & LoTSS quasar samples within LoLSS DR1 and with FIRST/LoLSS upper limits & 14,106 quasars\\
3.1 & $\alpha_{54,u}^{144} >0$ and $\alpha_{144,u}^{1400}<0$ & 103 MPS sources\\
\end{tblr}
\begin{flushleft}
\textit{Notes}: 

1. The 'u' in $\alpha_{54,u}^{144}$ and $\alpha_{144,u}^{1400}$ indicates the use of upper-limit flux densities.

2. The radio quasars in this work are SDSS quasars with FIRST counterparts within 2\arcsec, see Section \ref{o_r} for details.
\end{flushleft}
\end{table}

\subsection{Optical observation}
We use the Sloan Digital Sky Survey (SDSS) 16th quasar catalog \citep{2020ApJS..250....8L}, which covers a sky area of 14,000 $\rm deg^{2}$ and includes 750,414 quasars, making it the largest collection of quasars with spectroscopy confirmation to date. This catalog contains information on important parameters such as the redshift and optical magnitude. Additionally, \citet{Wu2022ApJS..263...42W} expand this catalog by providing supplementary quasar properties such as bolometric luminosities and SMBH masses.

We adopt $i-$band absolute magnitudes $k \text {-corrected}$ to redshift 2 ($M_i(z=2)$) from SDSS quasar catalog to obtain $i-$band absolute magnitudes $k \text {-corrected}$ to redshift 0 ($M_i(z=0)$, \citealt{2006AJ....131.2766R}) : 
\begin{equation}
M_i(z=0)=M_i(z=2)+2.5(1+\alpha) \log (1+z),
\end{equation}
where $\alpha$ and $z$ represent the optical spectral index and redshift, respectively. We take the canonical optical spectral index value of $\alpha = -0.5$ from \cite{2006AJ....131.2766R}.

\subsection{Radio observations}
The LoTSS survey is an ongoing LOFAR project that aims to provide a survey of the whole northern sky at 120-168 MHz. Here, we utilize the second data release 2 (DR2) from LoTSS (see \citealt{2022A&A...659A...1S} for details). This release comprises 4,396,228 radio sources and has a survey area of  5634 $\rm deg^{2}$, which corresponds to 27\% of the northern sky. At a resolution of $ 6''$ and with an average root mean square (rms) close to 100 $\mu$Jy/beam, it is the most sensitive, wide-area radio survey to date. This survey can observe extended sources with scales up to 1\degree \ \citep{2023A&A...678A.151H}. We use the catalog from \cite{2023A&A...678A.151H} which not only provides the flux densities of these extended sources in LoTSS DR2 but also offers information on their optical counterparts. This catalog includes 4,116,934 radio sources, of which 85\% have optical or infrared counterparts. 

LoLSS observes the northern sky with the declination > 24\degree \ using the Low Band Antenna across the 42-66 MHz frequency range
 (LBA, see \citealt{2023A&A...673A.165D} for details). The first data release (DR1) provides average flux densities for 42,463 radio sources at 54 MHz, with a resolution of $15''$.  This dataset encompasses 659 $\rm deg^{2}$ in the Hobby-Eberly Telescope Dark Energy Experiment (HETDEX) field \citep{2021ApJ...923..217G}. The rms noise level of the survey is approximately 1 mJy/beam. This survey facilitates the study of radio spectra at ultra-low frequencies with unprecedented sensitivity and resolution.  

Additionally, the Faint Images of the Radio Sky at Twenty-cm (FIRST, \citealt{1994ASPC...61..165B,1995ApJ...450..559B}) survey systematically maps the sky over 10,000 square degrees in the North and South Galactic Gap at 1.4 GHz using the NRAO Very Large Array (VLA). The survey achieves a resolution of $5''$ and has radio observations for 946,432 sources with a rms of 0.15 mJy \citep{2015ApJ...801...26H}. The coverage of the FIRST survey significantly overlaps with that of SDSS, providing a robust dataset for multiwavelength studies.

\subsection{Sample Selection}\label{sample_selection}


Our goal is to establish the largest quasar sample with multiband radio data. As mentioned earlier, the depth and coverage of radio data across different bands vary significantly. The shallow and small sky area observations can greatly limit the number of sources in the sample. For example, LoLSS DR1 (1-2 mJy/beam) and FIRST (0.15 mJy) data are less sensitive than LoTSS DR2 (100 $\mu$Jy/beam). 
Therefore, some faint sources are only detectable with LoTSS observations. Additionally, the sky coverage of LoLSS DR1 is quite limited, meaning that our study sample would only be constrained to that region of the sky.

We first selected sources primarily based on LoTSS, which is the largest and deepest survey. We then categorized the sample into two groups: one with detection and one with upper-limit data, depending on the availability of LoLSS or FIRST detections in the LoLSS DR1 sky coverage.

As our first step in sample selection, we cross-matched the LoTSS DR2 catalog containing value-added information (i.e, optical ids) with the SDSS 16th quasar catalog \citep{2023A&A...678A.151H,2020ApJS..250....8L}, using a matching radius of $5''$. This radius was chosen following \cite{2019A&A...622A..11G}, considering the resolution of the LOFAR map. We used the LoTSS DR2 optical counterpart catalog \citep{2023A&A...678A.151H} rather than the original DR2 radio catalog, as it provides accurate flux density measurements for extended sources, which is crucial for this study. From this process, we obtained 64,464 quasar samples from LoTSS.

Next, we cross-matched these quasars with the LoLSS DR1 \citep{2023A&A...673A.165D} and FIRST catalogs \citep{1994AAS...185.0803H,2015ApJ...801...26H} to collect multifrequency radio data and construct their radio spectra. We maintained the $5''$ radius for LoLSS DR1 to minimize contamination and preserve the reliability of the sources. Even though the SDSS 16th quasar catalog provides cross-matching results with FIRST, it does not supply the total flux density needed for our analysis. Therefore, we repeated the cross-matching with FIRST using a radius of $2 \arcsec$ for consistency \citep{2020ApJS..250....8L}. As a result, we identified 1,189 quasars with detections at 54 MHz from LoLSS, 144 MHz from LoTSS, and 1.4 GHz from FIRST. After excluding sources with measurement issues ($M_i(z=2)>0$), 1,187 quasars remained for further analysis. A summary of the sample selection process is provided in Table \ref{table1}.

To obtain meaningful upper limits for non-detected quasars, we applied the following method. For sources that were undetected in LoLSS but detected in LoTSS, we used the rms maps from \cite{2023A&A...673A.165D} and calculated the upper-limit flux densities. For FIRST data, given that the variability of the rms map is only about $15\%$, we adopted a fixed rms value (0.15 mJy from \citealt{2015ApJ...801...26H}) for all undetected sources. Upper-limit flux densities were defined as five times the rms noise. We used this factor of 5 to maintain consistency across different observations, as faint sources in LoTSS are generally cataloged at levels above 5 times the noise level \citep{2019A&A...622A...1S}. This process added 14,106 more quasars to our sample with multiwavelength radio data but with LoLSS or FIRST upper-limit flux densities in the LoLSS DR1 sky coverage.

Due to the relatively low resolution of LoLSS ($15''$), which could potentially merge emissions from two separate sources in LoTSS and lead to inaccurate flux measurements, we checked for any LoTSS sources within $15''$ of our quasar samples. After this check, we found no such cases in our sample, ensuring that none of the sources in our study were affected by this issue.

\subsubsection{Identifying extended sources}

In our selection of extended sources, we followed the method from \cite{2022A&A...659A...1S}, who distinguish between extended sources and point sources based on the ratio between the total ($S_{\rm total}$) and peak flux density ($S_{\rm peak}$) and the signal-to-noise ratio ($S/N$). Theoretically, the ratio between the total and peak flux density ($S_{\rm total}/S_{\rm peak}$) for point sources should be close to 1. However, due to imperfect calibration, this value can deviate. \cite{2022A&A...659A...1S} initially identified the distribution of isolated sources in terms of the $S_{\rm total}/S_{\rm peak}$ across different $S/N$. A threshold was established to include 99.9\% of point sources, which served as the demarcation line for distinguishing between point sources and extended sources. The relationship identified is:
\begin{equation}
    R_{99.9}=0.42+\left(\frac{1.08}{1+\left(\frac{S/N}{96.57}\right)^{2.49}}\right),
\end{equation}
where $R = \rm ln(\frac{S_{\rm total}}{S_{\rm peak}})$ and $S/N$ is defined as $\frac{S_{\rm total}}{\sigma_{total}}$. Using this method, we identified a total of 232 extended sources from a pool of 1,187 sources. This approach allows us to differentiate extended sources based on their flux characteristics, ensuring the scientific robustness of our sample selection. 

Since \cite{2023A&A...678A.151H} had already provided the total flux for extended sources, which is not available for the FIRST survey, we conducted a similar analysis for quasars detected in FIRST by calculating the total flux densities of multiple components at 1.4 GHz. For resolved sources in LoTSS DR2 with corresponding FIRST counterparts, we searched for additional counterparts within a $30''$ radius of the central position. We identified 174 resolved sources that had multiple FIRST counterparts. We then combined the flux densities of these multicomponent sources to obtain accurate total flux values for each quasar. These 174 sources are unresolved in LoLSS, so no additional flux correction is required.



\subsubsection{Optical and Radio selected quasars} \label{o_r}
SDSS selects potential quasar targets using both optical and radio selection methods \citep{2015ApJS..221...27M,2019A&A...622A..11G,2020ApJS..250....8L}. In the optical, the quasar candidates are identified based on their location in the color-color diagram, ensuring differentiation from stars. Additionally, the quasar candidates are cross-matched with the FIRST catalog to get a sample of the radio-selected quasars. Following the method from \cite{2019A&A...622A..11G}, we separate these two quasar samples by classifying objects with radio counterparts in FIRST as radio quasars, while those without are labeled as optical quasars, and obtain 1037 optical quasars and 150 radio quasars, as summarized in Table \ref{table1}.

\subsection{Identification of MPS sources}

As we mentioned above, PS sources are sources with peaked radio spectra. We follow the method from \cite{2017ApJ...836..174C} and \cite{2022A&A...668A.186S} to find the sources that peak around 144 MHz, namely MPS sources.  The spectral index is defined as
\begin{equation}
\alpha_{\nu_{1}}^{\nu_{2}}=\frac{\log \left(F_{\nu_{1}} / F_{\nu_{2}}\right)}{\log (\nu_{1}/ \nu_{2})},
\end{equation}
where $\alpha_{\nu_{1}}^{\nu_{2}}$ is the spectral index between 2 frequencies $\nu_{1}$ and $\nu_{2}$. The $F_{\nu_{1}}$ and $F_{\nu_{2}}$ are the flux densities at these frequencies. The criteria for identifying MPS sources are $\rm \alpha^{144}_{54} >=0.1$ and $\rm \alpha^{1400}_{144}<0$, where $\rm \alpha^{144}_{54}$ is the spectral index between the flux density of LoTSS 144 MHz and LoLSS 54 MHz and $\rm \alpha^{1400}_{144}$ is the spectral index between the flux density of FIRST 1.4 GHz and LoTSS 144 MHz.  We chose $0.1$ instead of $0.0$ for $\rm \alpha^{144}_{54}$ to exclude flat radio spectra (see, e.g., \citealt{2017ApJ...836..174C, 2022A&A...668A.186S}). Applying these criteria, we identified 61 MPS sources with three flux measurements. Additionally, we found 34 detected sources with $\rm 0 < \alpha^{144}_{54} < 0.1$ and $\rm \alpha^{1400}_{144} < 0$, which are flat-spectrum sources and are not considered in this work. From the sample with upper-limit flux densities from LoLSS/FIRST (see Sect. \ref{sample_selection}), we identified 103 additional MPS sources.

Since MPS sources typically have a dominant compact component, additional criteria were introduced in \cite{2022A&A...668A.186S} and \cite{2024A&A...689A.264B} to enhance the purity of their sample. These criteria included requirements for sources to be single-Gaussian and sufficiently isolated to prevent contamination from nearby sources. Specifically, they required that these sources should not be classified as "C" by the Python Blob Detection and Source Finder (PyBDSF, \citealt{2015ascl.soft02007M}), which indicates a single-Gaussian source within an island containing other Gaussian components. None of the 61 MPS sources in this study were categorized as "C," thus meeting this compactness criterion. Additionally, isolated sources were defined as having no neighboring sources within $45''$, based on the resolution of the radio data used in those studies. In our work, the lowest resolution is $15''$, based on LoLSS data, so we set our isolation criterion at $15''$. This ensures that two sources close enough to be distinguished in LoTSS will not be mistakenly identified as a single source in LoLSS. As we already conducted a similar check for all quasars in Section \ref{sample_selection}, the same applies to all the MPS sources: none of these sources have neighboring sources within a $15''$ range in LoTSS. Therefore, the 61 MPS sources we identified exhibit a clear turnover in their radio spectra, are compact and are not affected by flux from nearby sources.

\begin{figure}
   \centering
   \includegraphics[width=0.5\textwidth]{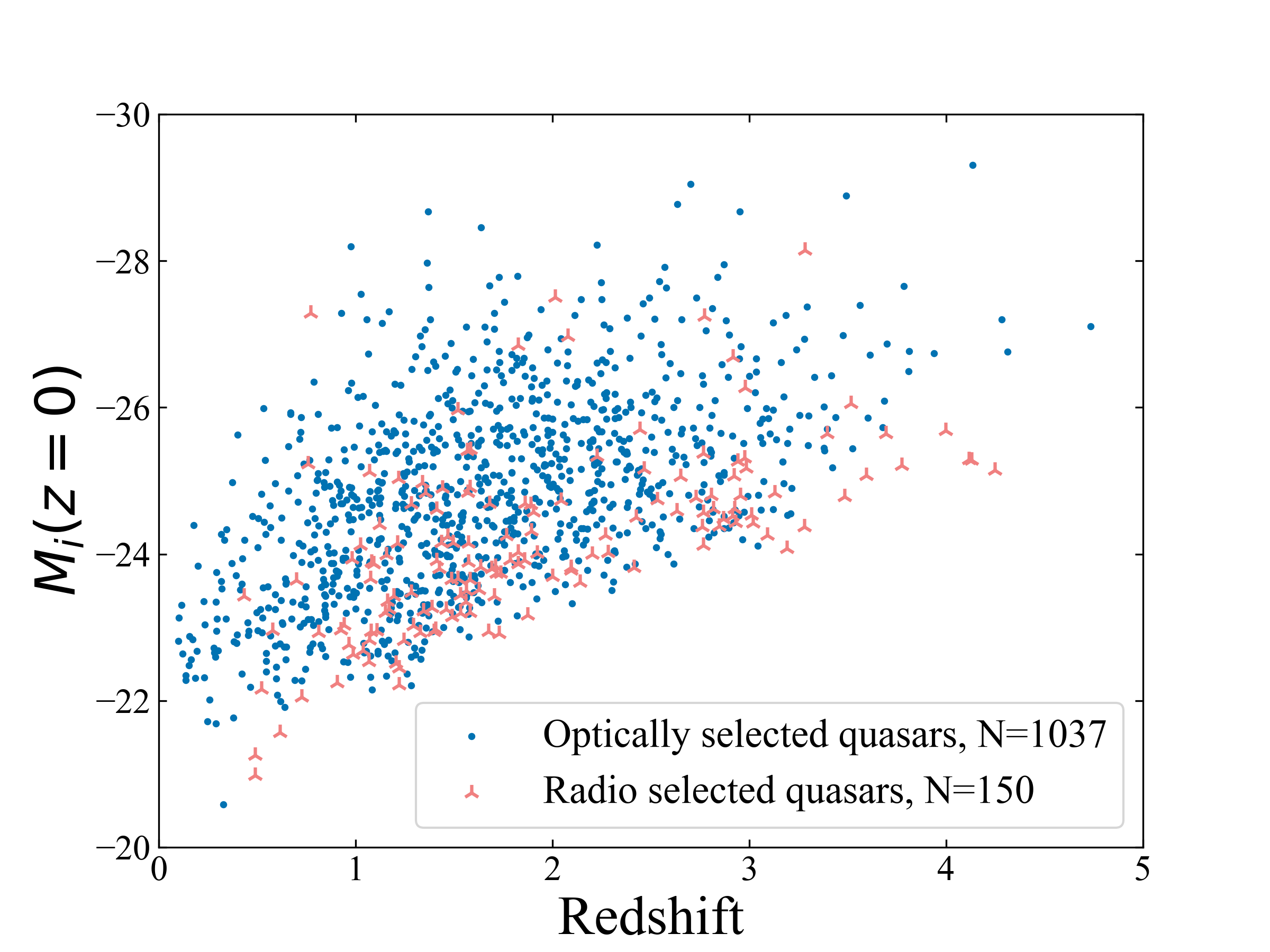}
   \caption{Distribution of $i-$band absolute magnitude as a function of redshift. The blue dots and red triangles represent the optically selected and radio selected quasars with the LOFAR detections, respectively. 
   }
              \label{i_z}%
    \end{figure}

\begin{figure*}
   \centering
   \includegraphics[width=\textwidth]{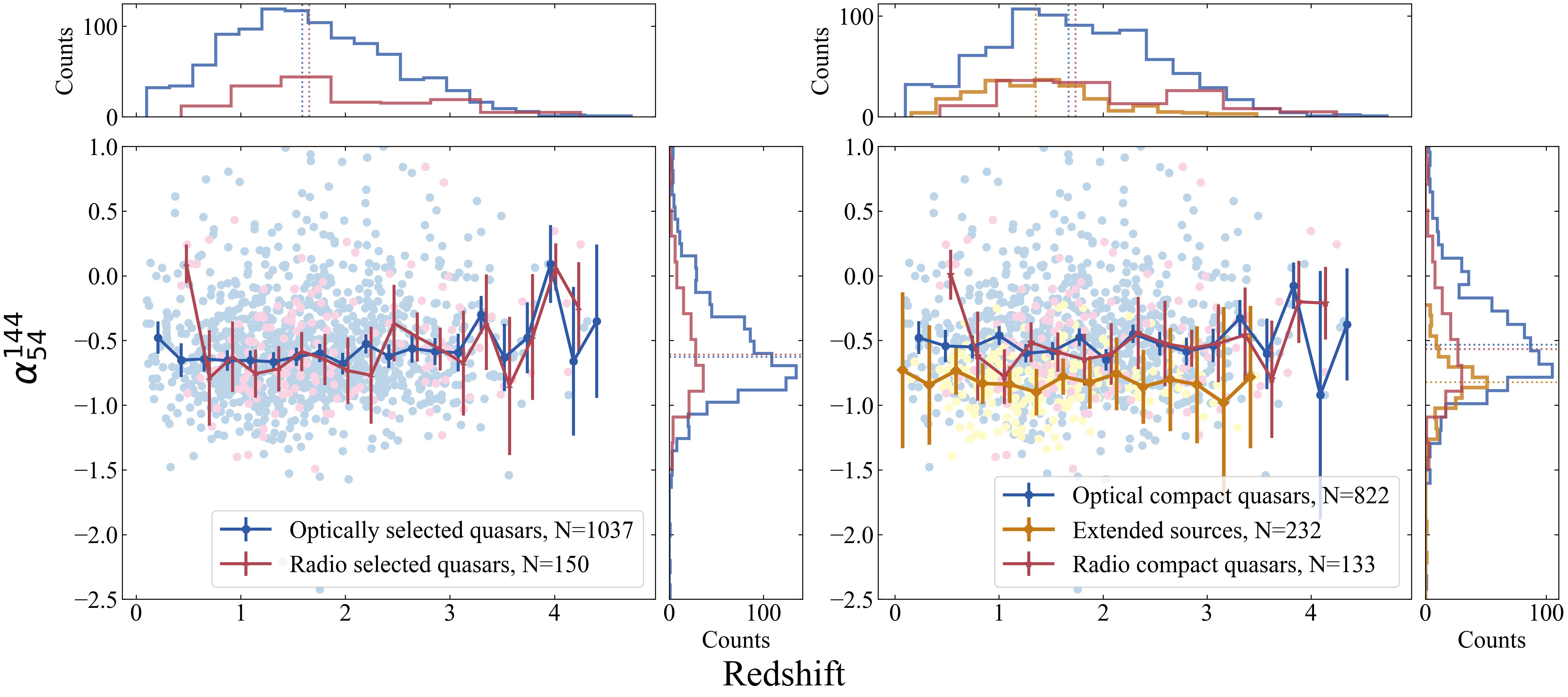}
   \caption{Relationship between the low-frequency spectral index ($\alpha^{144}_{54}$) and redshift. In the left panel, the distributions of optical-selected (blue) and radio-selected (red) quasars are shown, while the right panel shows the distributions for extended sources (yellow), optical compact sources (blue), and radio-compact sources (red). Error bars indicate the standard error of the mean ($\sqrt{\overline{x}/N}$) within each redshift bin. Blue, red, and orange lines represent optical-selected quasars, radio-selected quasars, and extended sources, respectively. Vertical lines mark the median values of each parameter in the corresponding histograms. Redshift bins are determined using Knuth's binning method to optimize bin width \citep{2006physics...5197K}. No significant trend is observed in the low-frequency spectral index ($\alpha^{144}_{54}$) with redshift for either sample.}
              \label{alpha_54_z}%
\end{figure*}

\begin{figure*}
   \centering
   \includegraphics[width=\textwidth]{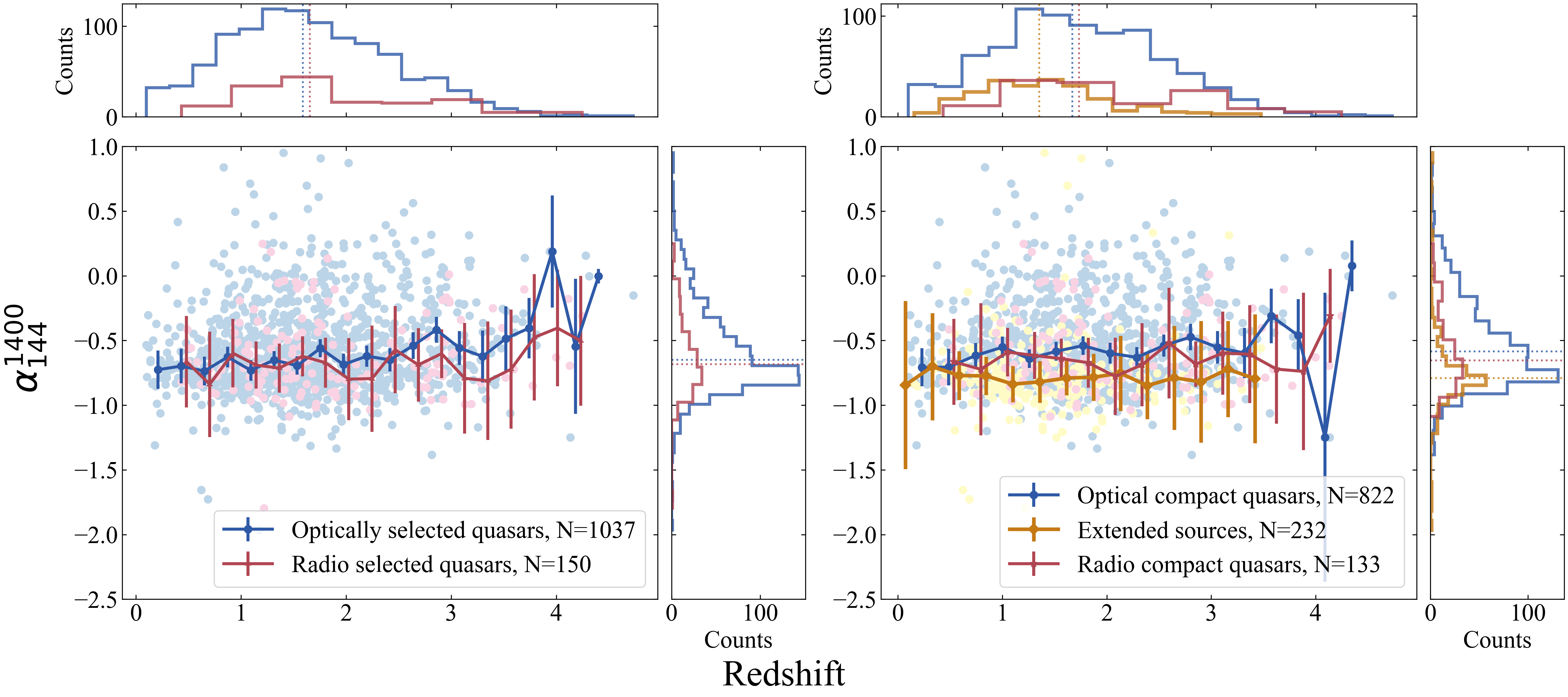}
   \caption{Relationship between the high frequency spectral index ($\alpha^{1400}_{144}$) and redshift. For a detailed explanation of the symbols, we refer to the caption of Fig. \ref{alpha_54_z}. No significant trend is observed in the high-frequency spectral index ($\alpha^{1400}_{144}$) with redshift for either sample.}
              \label{alpha_144_z}%
\end{figure*}

\begin{figure}
   \centering
   \includegraphics[width=0.5\textwidth]{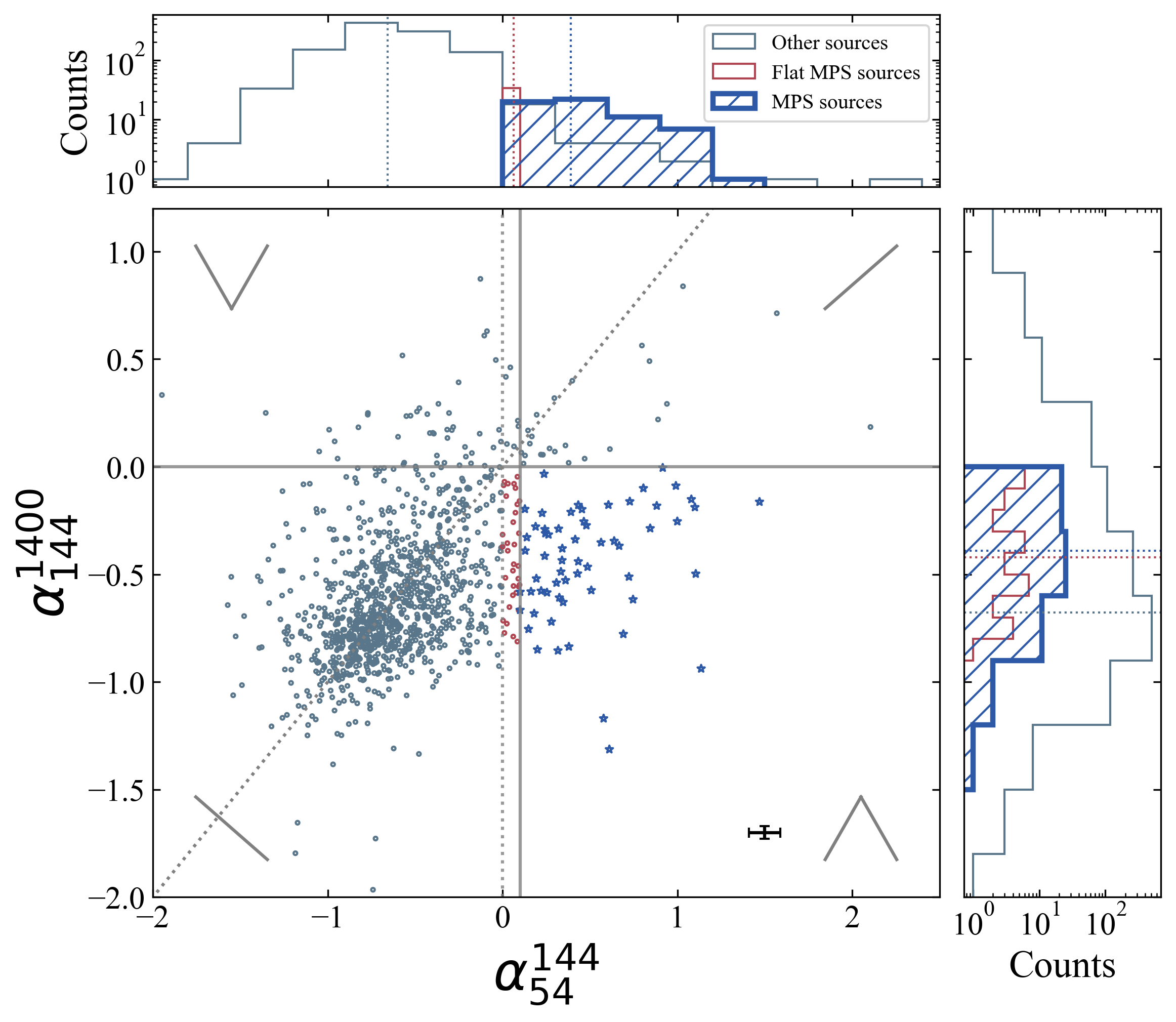}
   \caption{2D-Distribution of the spectral index for all quasars in the study. The four quadrants represent different shapes of the radio spectra. The blue star-shaped markers in the lower right quadrant are the MPS samples discussed in this paper. Red dots indicate sources that meet $\rm 0<\alpha^{144}_{54} < 0.1$ and $\rm \alpha^{1400}_{144}<0$ criteria, and light blue dots represent the remaining samples. The gray horizon, vertical, vertical dashed, and lines represent the $\alpha^{1400}_{144}=0$, $\alpha^{144}_{54}=0.1$, $\alpha^{144}_{54}=0$ and 1-to-1 lines. The typical median error of samples is represented by black lines in the bottom-right corner. The median errors are $0.09$ and $ 0.03$ for $\alpha^{144}_{54}$ and $\alpha^{1400}_{144}$, respectively. \textit{Top} and \textit{right}: Histogram of the spectral index, with colors corresponding to those in the middle panel. The vertical lines represent the median values of each sample.}
              \label{double_alpha}%
\end{figure}

\begin{figure*}
		
		\centering
		\begin{multicols}{3}
		    \includegraphics[width=0.333\textwidth]{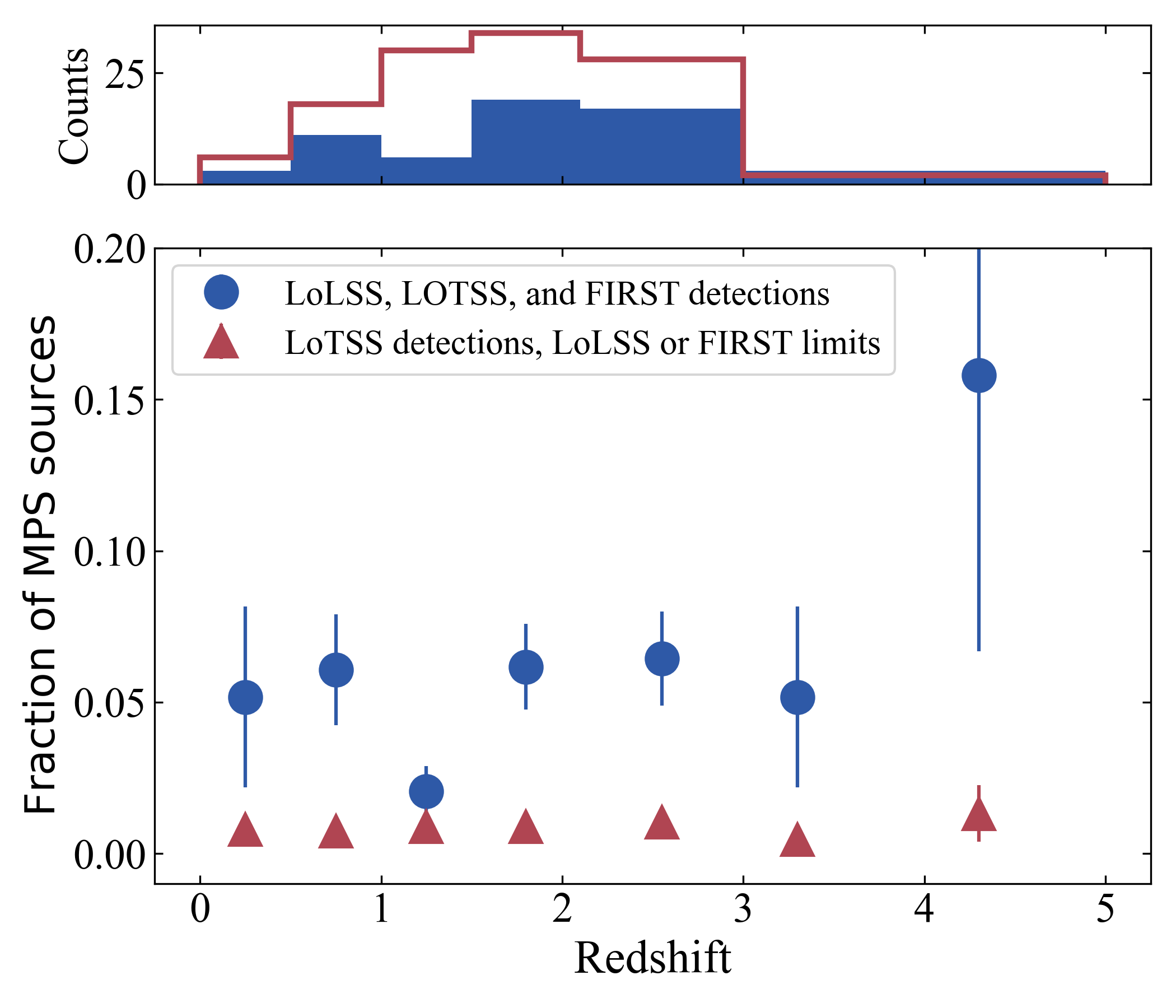}\par
			\includegraphics[width=0.333\textwidth]{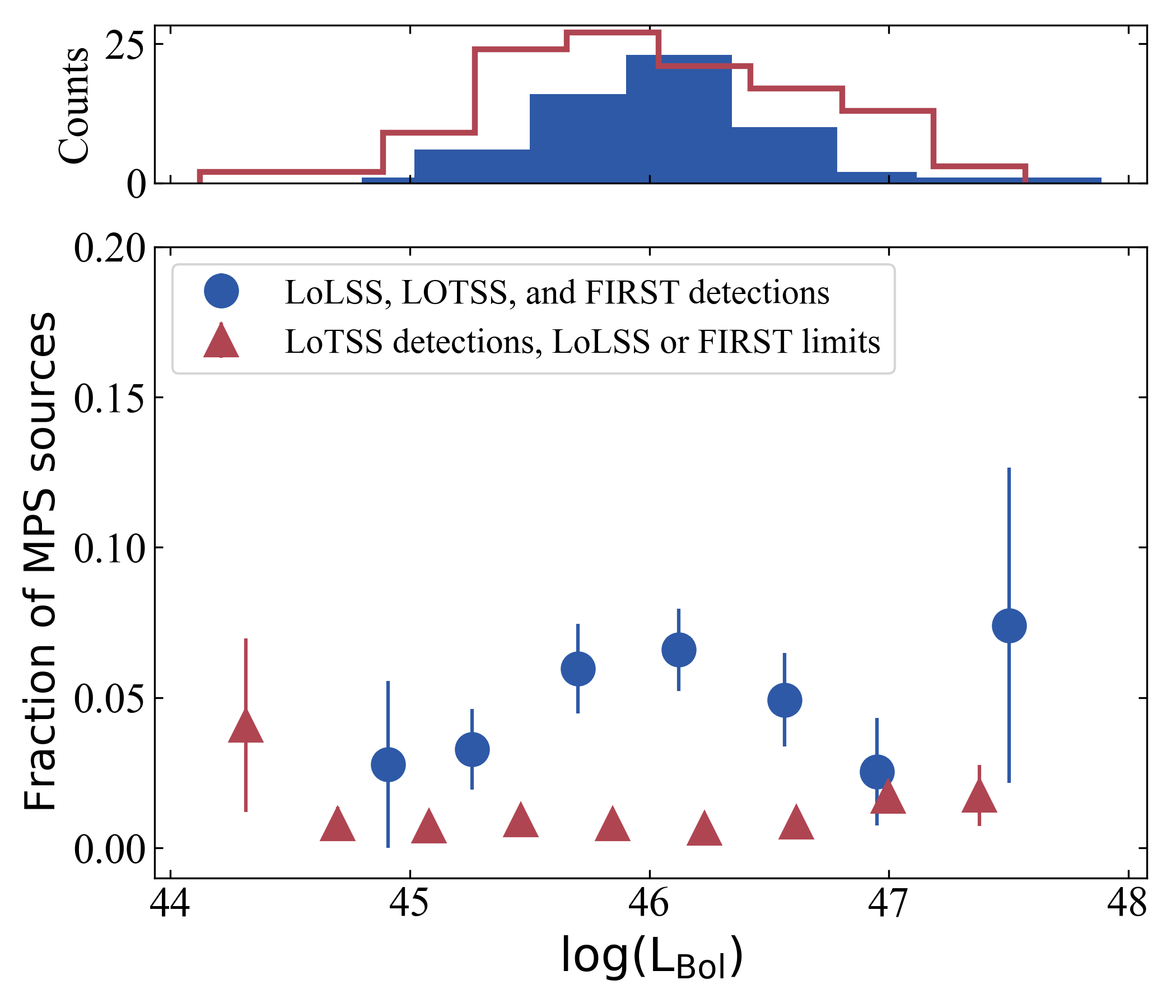}\par
			\includegraphics[width=0.333\textwidth]{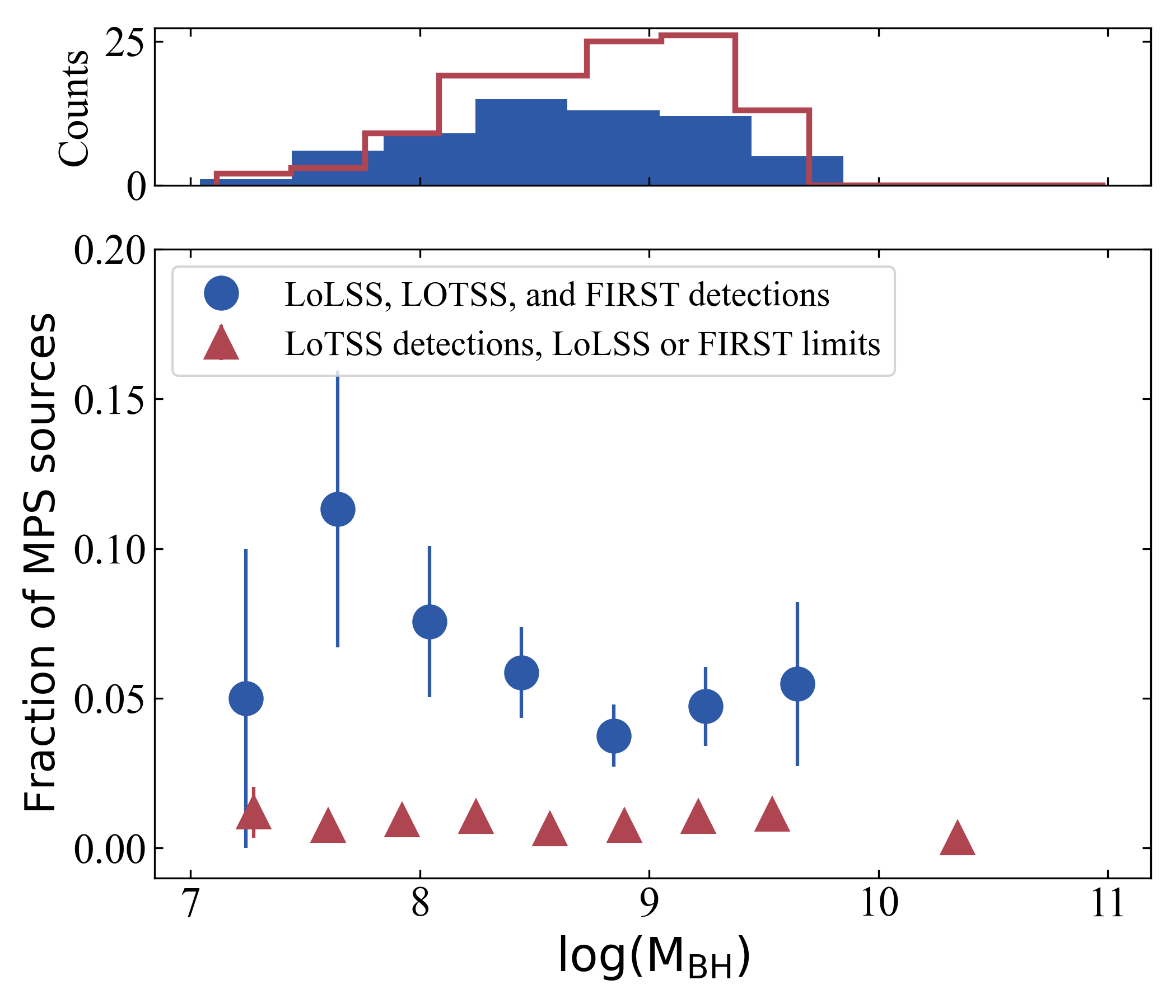}\par
			
		\end{multicols}

		\caption{From the left to the right panels, the relationships are shown between the fraction of MPS sources and redshift, bolometric luminosity ($L_{\text{Bol}}$), and black hole mass ($M_{\text{BH}}$). The blue dots represent sources detected by LoTSS, LoLSS, and FIRST, while the red triangles represent sources detected by LoTSS with LoLSS/FIRST upper limits in the  LoLSS DR1 sky coverage.}
		\label{fpeak_z}
	\end{figure*}

\begin{figure}
   \centering
   \includegraphics[width=0.5\textwidth]{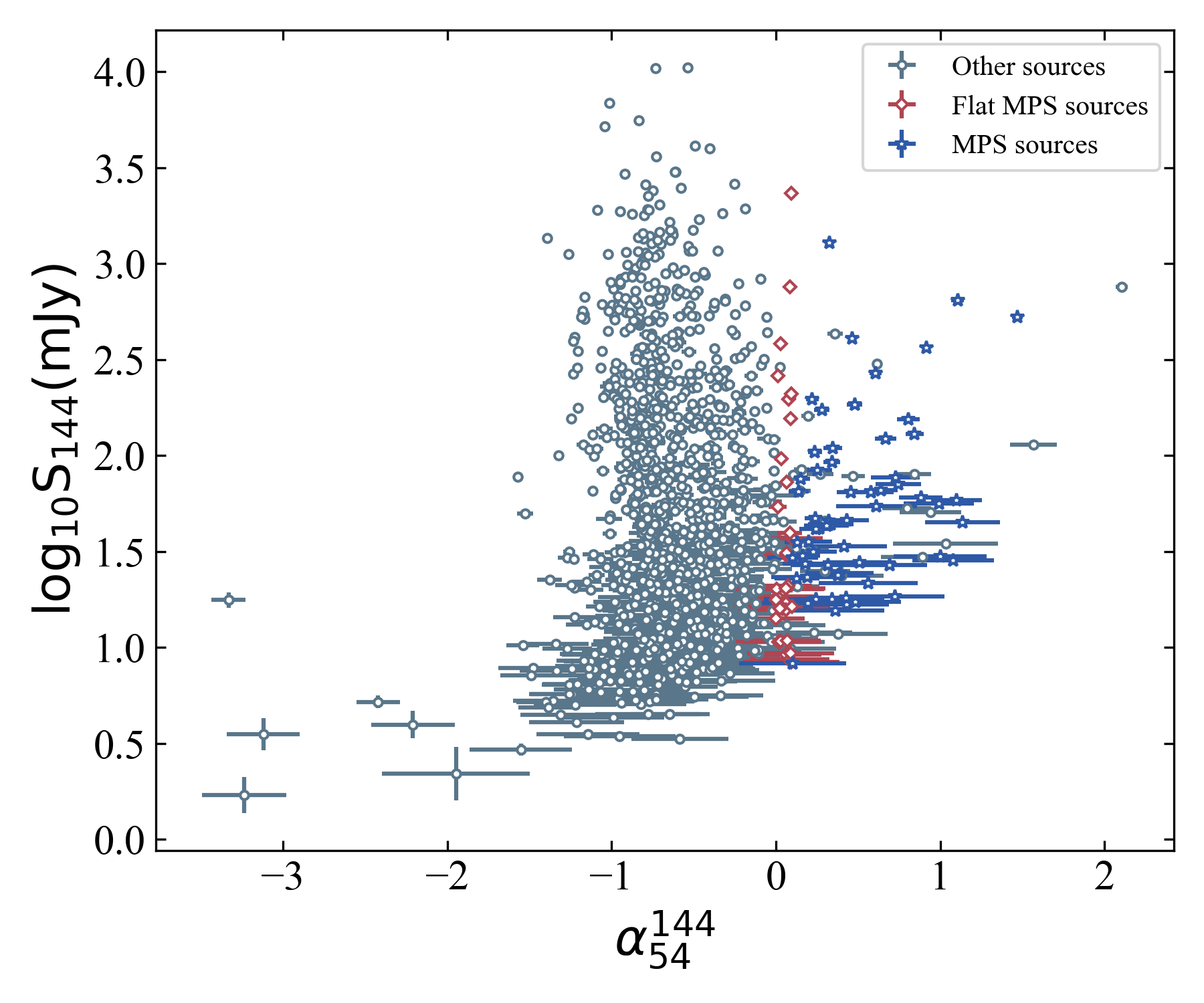}
   \caption{Relationship between the spectral index and flux density at 144 MHz for quasars in this work. The blue star-shaped markers are the MPS sources discussed in this paper. Red diamonds indicate sources with $\rm 0<\alpha^{144}_{54} < 0.1$ and $\rm \alpha^{1400}_{144}<0$ and light blue dots represent the remaining samples.}
              \label{flux_alpha}%
\end{figure}

\section{Results}

Fig. \ref{i_z} shows the $i-$band absolute magnitudes as a function of redshift for the LOFAR detected optical and radio selected quasars. Our data is distributed with i-band absolute magnitudes $<-20$, and the redshift spans the range from the local universe to $z\sim 5$. 

\subsection{The $\alpha-z$ relationship of quasars}\label{3_1}

In Fig. \ref{alpha_54_z} and Fig. \ref{alpha_144_z}, we present the redshift evolution of the spectral index. The number density of our sample peaks between redshift 1 and 2, as illustrated in the top panel of each figure.

The spectral index $\alpha^{144}_{54}$ as a function of redshift is shown in the left panel of Fig. \ref{alpha_54_z}. Both the optical-selected and radio-selected quasars show no trend with redshift. Additionally, the overall spectral index ($\alpha^{144}_{54}$) distribution remains constant with redshift, with median values being $-0.626\pm0.004$ for optically selected quasars and $-0.608\pm0.011$ for radio-selected quasars. These errors were calculated using error propagation \citep{Taylor1982}. The median spectral index appears to increase slightly toward higher redshifts, however, due to the limited sample size around a redshift of 4, a solid conclusion cannot be drawn. In the right panel of Fig. \ref{alpha_54_z}, we separate the extended sources from compact sources. Overall, the median spectral index ($\alpha^{144}_{54}$) of extended sources is steeper than that of compact sources. This difference is not significantly influenced by the potential loss of diffuse emissions, as LOFAR's extensive short baselines ensure excellent $uv$-coverage, enabling the recovery of such emission \citep{Hoang2018MNRAS.478.2218H, Botteon2020ApJ...897...93B, 2022A&A...659A...1S}. At low redshift, the median spectral index of radio compact sources is steeper than that of optical compact sources. Specifically, the median spectral index values for extended, optical compact, and radio compact sources are $-0.823\pm0.012$, $-0.532\pm0.011$, and $-0.567\pm0.030$, respectively. 

In Fig. \ref{alpha_144_z}, for both samples, the results for the high $\nu$ spectral index ($\alpha^{1400}_{144}$) are generally consistent with those for the low $\nu$ spectral index ($\alpha^{144}_{54}$). For optical and radio quasars, the median spectral indices are $-0.650\pm0.001$ and $-0.683\pm0.002$, respectively. The median spectral index of extended sources, with a value of $-0.790\pm0.006$, is steeper than that of compact sources: $-0.585\pm0.004$ for optical compact sources and $-0.655\pm0.011$ for radio compact sources. The median high $\nu$ spectral index ($\alpha^{1400}_{144}$) is steeper than the low $\nu$ spectral index ($\alpha^{144}_{54}$) for compact sources. This indicates that the overall radio spectrum of compact sources is flatter at lower frequencies compared to higher frequencies. 


\subsection{The evolution of MPS sources}
In Fig. \ref{double_alpha}, we present the low $\nu$ ($\alpha^{144}_{54}$) and high $\nu$ ($\alpha^{1400}_{144}$) spectral index radio color-color diagram. Most sources (around 85.2\%) exhibit a negative power-law spectral index in the bottom-left quadrant, consistent with previous work \citep{2017ApJ...836..174C}. The bottom-right quadrant shows the distribution of MPS sources (around 5.1\%). The top-right quadrant shows the GPS, and HPS sources (around 2.5\%). The top-left quadrant shows the sources with an upturn in their radio spectra (around 4.3\%). For the MPS sources, the median $\alpha^{144}_{54}$ and $\alpha^{1400}_{144}$ are $0.391\pm0.049$ and $-0.390\pm0.013$, respectively. 

The fraction of MPS sources ($\rm f_{MPS}$) as a function of redshift is shown in the left panel of Fig. \ref{fpeak_z}. The main plot shows that for detected sources in the three surveys, their fraction exhibits a peak around redshift 0.5 and 2. This aligns with the peaks around 2 in the distribution of the number of sources as a function of redshift in the subplot. We find that the fraction of MPS sources does not show a clear relationship with redshift, and the same result holds for samples with LoLSS or FIRST flux upper limits in the LoLSS DR1 sky coverage. The high fraction of MPS sources observed at $z>4$ is due to the sample number statistics rather than a genuine increase. Additionally, the number of MPS sources drops at $z>3$ which is related to the incomplete quasar identification at high redshift and the limited sky coverage of the survey \citep{Wu2022ApJS..263...42W,2023A&A...673A.165D}. In summary, the majority of MPS sources are indeed concentrated between redshift 2 and 3. However, the fraction of MPS sources does not change with redshift.


In the middle panel of Fig. \ref{fpeak_z}, we show that the fraction of MPS sources ($\rm f_{MPS}$) does not depend on the bolometric luminosity ($L_{\text{Bol}}$), which is derived from the rest-frame optical continuum luminosity \citep{Wu2022ApJS..263...42W}. Although there is an increase in the last luminosity bin, this is likely due to the limited sample size, as indicated by the large error bars. Therefore, we conclude that the fraction of MPS sources does not significantly change with bolometric luminosity. Since bolometric luminosity ($L_{\text{Bol}}$) is related to the properties of SMBHs, such as mass, spin, and accretion rate \citep{2019A&A...622A..11G, 2002apa..book.....F}, this result suggests that the activities of SMBHs may not strongly influence the formation of MPS sources. 

To further investigate the influence of SMBHs, the right panel of Fig. \ref{fpeak_z} presents the fraction of MPS sources ($\rm f_{MPS}$) as a function of SMBH mass. These black hole masses, estimated from measurements of the continuum and broad line emissions \citep{Wu2022ApJS..263...42W}, are concentrated between $10^{8.5}$ and $10^{9.5} \, \rm M_{\odot}$. Overall, the fraction of MPS sources remains consistent across different SMBH masses.

\section{Discussion}


\subsection{The constant $\alpha-z$ relationship of quasars\label{a-zdiscussion}}

In section \ref{3_1}, for quasars, we found that both the median $\alpha^{144}_{54}$ and $\alpha^{1400}_{144}$ spectral indices are steeper for extended sources compared to compact sources, which is consistent with previous work. Specifically, for $\alpha^{1400}_{144}$, the median value for extended sources is around -0.8 \citep{2018MNRAS.474.5008D,2019A&A...622A..11G}. For compact sources, studies generally found that the overall spectral index is flatter compared to extended sources. However, the specific results varied across different studies. For instance, in \cite{2019A&A...622A..11G}, the median values of $\alpha^{1400}_{144}$ for optical compact and radio compact sources are -0.26±0.02 and -0.36±0.06, respectively. Similarly, in \cite{2018MNRAS.474.5008D}, the spectral index ($\alpha^{1400}_{147}$) for compact sources is -0.5.

The difference in spectral indices between extended and compact sources is believed to be related to the emission originating from different parts of the sources \citep{1995PASP..107..803U}. Flat-spectrum compact sources are likely from core-dominated AGN, while steep-spectrum extended sources are likely from lobe-dominated regions \citep{1999AJ....117..677B}. The radio spectrum of AGN cores is typically flat (around $-0.5$), whereas the lobes have a steeper spectrum due to synchrotron and inverse Compton aging, also combined with adiabatic expansion \citep{1999AJ....117..677B,2018MNRAS.474.5008D}. 

We also found that the spectral indices ($\alpha^{144}_{54}$ and $\alpha^{1400}_{144}$) of quasars do not significantly change with redshift across $0<z<5$. This result is consistent with the result from \cite{2014MNRAS.443.2590S}, who found no clear steepening of $\alpha^{324}_{1400}$ for AGNs in the VLA-COSMOS field. Similar results were found for star-forming galaxies ($\alpha^{1400}_{610}$; \citealt{2010MNRAS.402..245I, 2014MNRAS.442..577T, 2015A&A...573A..45M}). \cite{2012MNRAS.420.2644K} also show a weak correlation between these two parameters in complete radio samples with spectral index at different frequencies (e.g., $\alpha^{1400}_{151}, \alpha^{750}_{151}$). 

However, previous results found a steepening of spectral indices for radio galaxies at higher redshift \citep{2000A&AS..143..303D,2002AJ....123..637D,2006MNRAS.371..852K,2014A&A...569A..52S}. This steepening could be attributed to the K-correction of the concave spectrum, increasing environmental density at higher redshift, or secondary effects of the luminosity-spectral index relationship \citep{1990ApJ...363...21C,1998A&A...329..809A,2008A&ARv..15...67M}. Since we observed no steepening of the spectral index at lower frequencies for quasars, these mechanisms might not be important for the spectral evolution of quasars. First, the K-correction effect, which causes higher redshift radio galaxies to appear steeper due to a shift in the observed spectral, may not apply to quasars due to their typically flatter radio spectra \citep{Hutchings1987, Tadhunter2016, Gloudemans2023}. Secondly, the environmental density that could cause radio spectral steepening in radio galaxies is likely less impactful for quasars. This is supported by the positive relationship between jet power and radio luminosity \citep{Cavagnolo2010ApJ...720.1066C, Godfrey2016MNRAS.456.1172G}, and the evidence that quasars in this work have a powerful median radio luminosity of $\log_{10}(L_{144 \  \rm MHz}) = 26.83 \pm 0.86 \ \rm W/Hz$, calculated using data from \cite{2023A&A...678A.151H}. Such high radio luminosities indicate that these quasars possess powerful jets. These jets and high-energy emissions make quasars less sensitive to increased environmental density \citep{Lietzen2009A&A...501..145L, Lietzen2011A&A...535A..21L}. Lastly, the luminosity-spectral index relationship may also differ for quasars, which are often dominated by accretion-related emissions rather than lobe-dominated synchrotron emissions, making their spectral evolution less dependent on redshift-related environmental changes  \citep{Barthel1989ApJ...336..606B, 1995PASP..107..803U, Tadhunter2016}. These factors suggest that the steepening mechanisms observed in radio galaxies may not be important for the spectral evolution of quasars.


\subsection{The mechanism behind the MPS sources \label{MPSturnover}}

Two possible absorption scenarios have been proposed to explain the turnover of the radio spectrum. Synchrotron Self-Absorption (SSA) occurs when a source is optically thick, in the case of a uniform plasma, leading to a spectral index close to 2.5 at low frequencies \citep{1997AJ....113..148O}. Free-free absorption (FFA) is due to photons being absorbed by the ambient dense medium. If the absorption is within the source, the typical spectral index should be 2.1 \citep{2015ApJ...809..168C, 2019A&A...628A..56K}. The external absorption will make the low-frequency spectral index steeper \citep{1997ApJ...485..112B,2019A&A...628A..56K}.

In this work, we identified 61 MPS sources with peaks at the extremely low frequency of 144 MHz. We found that for these sources, the median $\alpha^{144}_{54}$ is 0.391 which does not allow us to draw definitive conclusions on the general absorption mechanism. The significant discrepancy between the observed spectral index (0.391) and the predicted value (2.1 or 2.5) could be due to our data points being too close to the peak frequency, or possibly because the peak frequency lies within the 54-144 MHz range. Also, the selection effect could impact the detection of MPS sources in this work. As shown in Fig. \ref{flux_alpha}, steep spectral indices (around 2) tend to have higher flux densities at 144MHz. Thus identifying the predicted steep spectral indices, such as 2.1 or 2.5, requires that sources are bright in LoTSS. This requirement might contribute to the absence of steep spectral indices observed here. Consequently, the additional data at lower frequencies, such as from the LOFAR Decameter Sky Survey (LoDeSS) with observations between 10-30 MHz \citep{2024NatAs...8..786G} or other data around 100 MHz, would allow us to perform spectral fitting to determine both the peak and the spectral index below the peak more accurately.

We also found the constant relationship between the fraction of MPS sources and redshift, bolometric luminosity, and SMBH mass. This suggests that the occurrence of MPS sources is relatively independent of cosmic evolution, radiation intensity, and black hole mass, pointing to a distinct physical mechanism that remains stable across different evolutionary stages and environments. This stability aligns with studies indicating that quasar and AGN spectral indices can vary independently of redshift and environmental conditions \citep{Falcke1995A&A...293..665F, Lietzen2009A&A...501..145L, Lietzen2011A&A...535A..21L, Tadhunter2016}.

\section{Summary}
The large quasar sample provided by SDSS and the high resolution, high sensitivity low-frequency radio data from LOFAR and VLA enable us to study the $\alpha-z$ relationship of quasars at frequencies below 100 MHz. We obtained a quasar sample of 1,187 sources with detection in SDSS, LoLSS, LoTSS, and FIRST surveys. We found constant spectral indices ($\alpha^{144}_{54}$ and $\alpha^{1400}_{144}$) as redshift increases (0<z<5). Additionally, we confirmed that the extended sources in our quasar sample have steeper spectra than compact sources at lower frequencies. This could be explained by the radio emission originating from different regions. The flat spectra would be from the compact core region, the emission of extended sources are mostly from the lobes. 

 By employing cuts for spectral indices $\alpha^{144}_{54}>=0.1$ and $\alpha^{1400}_{144}<0$, we identify 61 MPS sources. Our analysis reveals that the fractions of these sources remain constant with redshift, bolometric luminosity, and SMBH mass. These findings suggest that MPS sources are governed by unique physical conditions or formation mechanisms that remain stable across different evolutionary stages and environments. However, it is important to note that the small sample size constrains this analysis. Future work should focus on expanding the sample size and exploring a lower and broader frequency coverage to improve our understanding of the turnover model of MPS sources. Detailed spectral modeling could also help locate the turnover frequency and spectral indices below. Further analysis of the environment and host galaxy properties may also shed light on the external impact on the formation of MPS sources.


\begin{acknowledgements}
We thank the anonymous referee for their valuable comments. SZ thanks Jinyi Liu, Yuming Fu, and Joseph Callingham for their helpful assistance.
LOFAR data products were provided by the LOFAR Surveys Key Science project (LSKSP; https://lofar-surveys.org/) and were derived from observations with the International LOFAR Telescope (ILT). LOFAR \citep{2013A&A...556A...2V} is the Low Frequency Array designed and constructed by ASTRON. It has observing, data processing, and data storage facilities in several countries, which are owned by various parties (each with their own funding sources), and which are collectively operated by the ILT foundation under a joint scientific policy. The efforts of the LSKSP have benefited from funding from the European Research Council, NOVA, NWO, CNRS-INSU, the SURF Co-operative, the UK Science and Technology Funding Council and the Jülich Supercomputing Centre. SZ acknowledges support from the CSC (China Scholarship Council)-Leiden University joint scholarship program. FdG acknowledges the support of the ERC Consolidator Grant ULU 101086378.
\end{acknowledgements}

\bibliographystyle{aa} 
\bibliography{aa} 

\end{document}